\documentclass{article}
\usepackage[super,sort&compress,comma]{natbib} 
\usepackage[version=3]{mhchem}
\usepackage[left=1.5cm, right=1.5cm, top=1.785cm, bottom=2.0cm]{geometry}
\usepackage{balance}
\usepackage{times,mathptmx}
\usepackage{sectsty}
\usepackage{graphicx} 
\usepackage{lastpage}
\usepackage[format=plain,justification=justified,singlelinecheck=false,font={stretch=1.125,small,sf},labelfont=bf,labelsep=space]{caption}
\usepackage{float}
\usepackage{fancyhdr}
\usepackage{fnpos}
\usepackage{chngcntr} 
\usepackage[english]{babel}
\addto{\captionsenglish}{%
  
}
\usepackage{array}
\usepackage{droidsans}
\usepackage{charter}
\usepackage[T1]{fontenc}
\usepackage[usenames,dvipsnames]{xcolor}
\usepackage{setspace}
\usepackage[compact]{titlesec}
\usepackage{hyperref}

\usepackage{epstopdf}

\definecolor{cream}{RGB}{222,217,201}

\begin{document}



\makeFNbottom
\makeatletter
\renewcommand\LARGE{\@setfontsize\LARGE{15pt}{17}}
\renewcommand\Large{\@setfontsize\Large{12pt}{14}}
\renewcommand\large{\@setfontsize\large{10pt}{12}}
\renewcommand\footnotesize{\@setfontsize\footnotesize{7pt}{10}}
\makeatother

\renewcommand{\thefootnote}{\fnsymbol{footnote}}
\renewcommand\footnoterule{\vspace*{1pt}%
\color{cream}\hrule width 3.5in height 0.4pt \color{black}\vspace*{5pt}} 
\setcounter{secnumdepth}{5}

\makeatletter 
\renewcommand\@biblabel[1]{#1}            
\renewcommand\@makefntext[1]%
{\noindent\makebox[0pt][r]{\@thefnmark\,}#1}
\makeatother 
\renewcommand{\figurename}{\small{Fig.}~}
\sectionfont{\sffamily\Large}
\subsectionfont{\normalsize}
\subsubsectionfont{\bf}
\setstretch{1.125} 
\setlength{\skip\footins}{0.8cm}
\setlength{\footnotesep}{0.25cm}
\setlength{\jot}{10pt}
\titlespacing*{\section}{0pt}{4pt}{4pt}
\titlespacing*{\subsection}{0pt}{15pt}{1pt}

\fancyfoot{}
\fancyfoot[LO,RE]{\vspace{-7.1pt}\includegraphics[height=9pt]{head_foot/LF}}
\fancyfoot[CO]{\vspace{-7.1pt}\hspace{13.2cm}\includegraphics{head_foot/RF}}
\fancyfoot[CE]{\vspace{-7.2pt}\hspace{-14.2cm}\includegraphics{head_foot/RF}}
\fancyfoot[RO]{\footnotesize{\sffamily{1--\pageref{LastPage} ~\textbar  \hspace{2pt}\thepage}}}
\fancyfoot[LE]{\footnotesize{\sffamily{\thepage~\textbar\hspace{3.45cm} 1--\pageref{LastPage}}}}
\fancyhead{}
\renewcommand{\headrulewidth}{0pt} 
\renewcommand{\footrulewidth}{0pt}
\setlength{\arrayrulewidth}{1pt}
\setlength{\columnsep}{6.5mm}
\setlength\bibsep{1pt}

\makeatletter 
\newlength{\figrulesep} 
\setlength{\figrulesep}{0.5\textfloatsep} 

\newcommand{\topfigrule}{\vspace*{-1pt}%
\noindent{\color{cream}\rule[-\figrulesep]{\columnwidth}{1.5pt}} }

\newcommand{\botfigrule}{\vspace*{-2pt}%
\noindent{\color{cream}\rule[\figrulesep]{\columnwidth}{1.5pt}} }

\newcommand{\dblfigrule}{\vspace*{-1pt}%
\noindent{\color{cream}\rule[-\figrulesep]{\textwidth}{1.5pt}} }

\makeatother

\twocolumn[
  \begin{@twocolumnfalse}
\vspace{3cm}
\sffamily
\begin{tabular}{m{4.5cm} p{13.5cm} }

\includegraphics{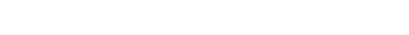} & \noindent\LARGE{\textbf{The role of extensional rheology in the oral phase of swallowing: an \textit{in vitro} study}} \\
\vspace{0.3cm} & \vspace{0.3cm} \\

 & \noindent\large{Marco Marconati \textit{$^{a}$} and Marco Ramaioli $^{\ast}$\textit{$^{a,}$}\textit{$^{b}$}} \\

\includegraphics{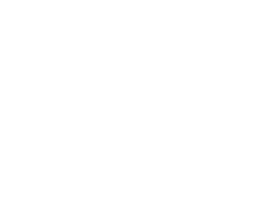} & \noindent\normalsize{Swallowing disorders deteriorate significantly the quality of life and can be life-threatening. Texture modification using shear thinning food thickeners have proven effective in the management of dysphagia. Some studies have recently considered the positive role of cohesiveness, but there is still an insufficient understanding of the effect of the rheological properties of the liquid bolus on the dynamics of bolus transport, particularly when elasticity and extensional properties are combined with a shear thinning behaviour. This study combines steady shear, SAOS and capillary breakage extensional rheometry with an \textit{in vitro} method to characterize the oral transport of elastic liquids. Bolus velocity and bolus length were measured from \textit{in vitro} experiments using image analysis and related to the shear and extensional properties. \textcolor{black}{A theory describing the bolus dynamics shows that the elastic and extensional properties do not influence significantly the oral transit dynamics.} Conversely, \textit{in vitro} results suggest that the extensional properties can affect the transition from the oral to the pharyngeal phase of swallowing, where thin, viscoelastic liquids lead to a fast transit, lower \textcolor{black}{oral} post-swallow residues and more compact bolus \textcolor{black}{with a smoother surface, which may suggest a lower risk of fragmentation}. This mechanistic explanation suggests that the benefit of the extensional properties of thin viscoelastic liquids in the management of dysphagia should be further evaluated in clinical trials.} \\
\end{tabular}

 \end{@twocolumnfalse} \vspace{0.6cm}

  ]

\renewcommand*\rmdefault{bch}\normalfont\upshape
\rmfamily
\section*{}
\vspace{-1cm}

\footnotetext{\textit{$^{a}$~Department of
Chemical and Process Engineering,
Faculty of Engineering and Physical
Sciences, University of Surrey,
Guildford, GU2 7XH, United Kingdom}}

\footnotetext{\textit{$^{c}$~UMR SayFood,
AgroParisTech, INRAE, Université Paris-Saclay,
1 avenue Lucien Brétignières,
78850 Thiverval Grignon,
France}}
\footnotetext{\textit{$^{\ast}$~Corresponding author. E-mail: marco.ramaioli@inrae.fr}}

\section{Introduction}

Swallowing disorders (dysphagia) are increasingly common, due to the ageing of the population. Dysphagia compromises strongly quality of life, concerns more than 40\% of residents in nursing homes \cite{Khan2014} and can cause life-threatening complications \cite{Leonard2013}.  

Clinical management of dysphagia requires a patient-centric approach considering a broad spectrum of compensatory techniques, such as a careful control of the sip size, the adjustment to the food texture and the correction to the eating posture \cite{Leonard2013}. Alterations to food texture are very common to promote safer swallowing in patients affected by mild to moderate forms of swallowing impairment \cite{Leonard2013,Steele2015a,Cichero2016}. Minced and pureed foods offer nutrition support for patients with dental and oral problems, impaired mastication and sarcopenia. Thickened beverages represent the primary means of hydration for many patients who are at risk of liquid aspiration.

A standardization of the rheological properties of thickened liquids based on their steady shear viscosity is attractive, but prone to oversimplification: the rheological behaviour of liquids can indeed be quite diverse, spanning from Newtonian sucrose solutions, such as syrups and honey, to highly shear thinning, viscoelastic and thixotropic fluids. Moreover the inclusion of finely ground tablets or microparticulate formulations \cite{MarconatiUCL}, or the inhomogeneity of the liquid carrier are all elements that may change the dynamics of swallowing. The impact of alterations of bolus viscosity to the perceived ease to swallow and mouthfeel has been the subject of several studies in literature, sometimes with contrasting results \cite{Clave2006,Cichero2016,Leonard2014,Steele2015a}.

A meaningful comparison of food thickeners with different shear thinning rheology requires defining the range of shear rates involved during oral processing and swallowing\cite{Chen2009,Steele2015a}. Although it is common to report viscosity at shear rates of 50 reciprocal seconds, the pharyngeal and esophageal phases are believed to generate higher shear rates, depending on the rheological properties of the swallowed bolus \cite{Nicosia2013,Nishinari2016,Newman2016, Gallegos2012}. \textcolor{black}{As a result some authors advocate for characterizing the shear rheology of texture modifiers also at 300 reciprocal seconds \cite{Gallegos2012}}.
The lack of general agreement on the nomenclature and testing protocols for texture-modified liquids has recently led to the establishment of the international dysphagia diet standardization initiative (IDDSI) that has introduced a new set of testing methods and nomenclature to characterize the consistency of foods and drinks \cite{Cichero2016a}. Under the IDDSI framework, thickened beverages are classified via a simple gravity discharge test \cite{Cichero2016a}. This flow test has already been considered by several authors \cite{Hadde2019,Barbon2018,Dantas2018,Ong2018}, but a relationship between the liquid rheology and the IDDSI scale has not yet been comprehensively investigated. 

While there is general agreement that texture modified liquids reduce aspiration, but the mechanistic reasons why this occurs are still debated. Vilardell \textit{et al.} \cite{Vilardell2016} investigated the effect of bolus viscosity on a significant base of 122 post-stroke patients, showing no significant kinematic differences with bolus consistency. Both starch and XG based thickeners were considered in this study and, per each sample, and three different levels of viscosity were used. Accordingly, the characteristic times of laryngeal elevation and airway closure did not show statistically significant variations. Also, the final velocity of the bolus in case of thin liquids (0.20-0.52 m/s) was not significantly modified at \textit{Nectar-thick} (0.18-0.5 m/s) \textit{nor spoon-thick} viscosity (0.19-0.47 m/s) \cite{Vilardell2016}. 

Using diagnostic Doppler ultrasounds Hasegawa \textit{et al.} measured the velocity in the oropharynx of boluses composed of water, yogurt, as well as gelatine and gellan gels and they registered average speeds of around 0.1 m/s for semisolid boluses, 0.2 m/s for fluid yogurt, and a maximum velocity of 0.5 m/s for water \cite{Hasegawa2005}.
Later, with a similar experimental apparatus, also Kumagai \textit{et al.} \cite{Kumagai2009} were able to show a noticeable variation in bolus velocity in the oropharynx as a function of bolus viscosity. They showed that that both the maximum velocity and the shape of the ultrasound Doppler echo considerably changed in respect of the liquid considered. Interestingly, it was shown that there was a logarithmic correlation between bolus maximum velocity and bolus viscosity measured at 10 reciprocal seconds \cite{Kumagai2009}. The study of Kumagai \textit{et al.} is however of limited applicability considering that the sampling population consisted of just a single healthy subject.

The study by Vilardell \textit{et al.} showed that the use of thicker solutions results in an appreciable reduction in the risk of penetration-aspiration, and also highlight the fact that thicker media leave more post swallow residues in the pharynx \cite{Vilardell2016}. The same conclusion was also clear from Steele \textit{et al.} who highlighted the fact that individuals affected by neurological diseases need to use a greater number of tongue actions to successfully swallow a \textit{pudding-thick} consistency, than for a thin liquid bolus, thereby suggesting that clearance was worse with the thicker consistency \cite{Steele2015a}. This was more evident when using starch-based thickened solutions, whilst a lower amount of oral and pharyngeal residues was observed after swallowing gum-based aqueous solutions \cite{Steele2015a}. 

The swallowing flow behaviour of more complex fluids has not yet received sufficient consideration. Whilst it is well accepted that both shear and extensional flows occur during swallowing \cite{Hadde2019,Newman2016,LaFuente2017,Salinas-V??zquez2014}, the importance of the elastic properties has only recently been investigated. The attribute of bolus cohesiveness, as a main contributor to triggering swallowing, was put in relation to the elongational properties of fluids \cite{Chen2009,Cichero2006}. However, the significance of this attribute has not yet been comprehensively investigated in dysphagic patients. Sensory tests comparing Newtonian and constant viscosity elastic fluids (Boger) on dysphagic subjects seem to suggest an improvement in the swallow-ability when introducing an elastic component. Additional studies are however needed to confirm these findings, as similar differences were not appreciated by the control group \cite{Waqas2017,Nystrom2015a}.
Nystrom \textit{et al.} stated that the superior performance of gum-based thickeners for dysphagia management, compared to starch based thickeners, can be directly related to the difference in terms of extensional viscosity \cite{Nystrom2015a}. At similar level of consistency, the authors found that commercial Xanthan gum (XG) based thickeners have a considerably higher extensional viscosity than starch based counterparts \cite{Nystrom2015a}. The same conclusion was also drawn by Hadde and Chen who studied the capillary break-up of commercial XG and starch-based thickeners \cite{Hadde2019}. The incorporation of saliva represents an additional degree of freedom that can impact on the bolus rheology \cite{Choi2014} owing to the peculiar extensional properties of saliva \cite{Wagner2017}.
A review was recently published summarising the evidence suggesting that extensional properties and cohesiveness could help preventing aspiration and summarizing many open research questions \cite{Nishinari2019}.  \textcolor{black}{A new \textit{in vivo} study on the impact of extensional properties on swallowing was published after the submission of the fisrt draft of this manuscript \cite{Hadde2019_2}. The authors considered only liquids with weak extensional properties, but claim nevertheless that ”higher extensional viscosity fluids reduced the elongation of the bolus during swallowing, thus potentially reducing the risk of post-swallow residue due to bolus breakage.”}

Several \textit{in vitro} models have been developed to understand the biomechanics of swallowing and the role of bolus rheology \cite{Marconati2018d}.
This study presents a full rheological characterization in shear and extension of different liquids and uses an \textit{in vitro} model, previously developed in \cite{Hayoun2015a,Mowlavi2016,Marconati2017b}, to clarify the role of extensional properties during the oral phase of swallowing.

\section{Materials and Methods}
This study considers three different model liquid boli. An aqueous solution of a commercial Xanthan gum based thickener (Resource\textsuperscript{\textregistered} ThickenUp\texttrademark Clear, Nestl\'e Health Science), in the following TUC, was chosen to provide a reference relevant to commercial products used in the management of dysphagia. \textit{Nectar-thick} solutions of TUC were prepared adding 100 mL of deionized water to 1.2 g of TUC (1.19 \% w/w). 
\textcolor{black}{The investigation of the effect of elasticity was instead carried out using aqueous solutions of polyethylene oxide (PEO, CAS 25322-68-3, average molecular weight $M_W=10^6$ g/mol), a polymer commonly used as a binder in pharmaceutical and cosmetics. Dilute aqueous solutions of PEO have been studied extensively for their extensional properties \cite{Arnolds2010} and represent a stable system with limited rheological complexity, compared to food systems.}
Aqueous solutions with concentrations ranging from 2 to 5\% w/w PEO were prepared in deionized water with the polymer left hydrating overnight in sealed containers under magnetic stirring. 

Aside from TUC and PEO, this study considered also cereal extract samples provided by Nestl\'e Research (Lausanne, CH). The total dry matter of the polysaccaride solutions was approximately 0.3 and 1 \% w/w for the low and high concentration samples respectively. 
The frozen cereal extract samples were thawed in a refrigerator at 4$^{\circ}$C for 12 hours, then left to equilibrate at ambient temperature for 4 hours, prior to the rheological characterization and \textit{in vitro} tests.

\subsection{Rheological characterization}
The rheological measurements included tests in steady shear, small amplitude oscillatory shear (SAOS) and extensional flow.

Measurements of shear viscosity were taken in triplicates, at 22$^{\circ}$C, with a Paar Physica UDS 200 controlled stress rheometer (Anton Paar GmbH, Graz, Austria). The flow curves were obtained in rate range of shear rates between 0.1 to 500 reciprocal seconds using a cone and plate geometry (d=75 mm $\alpha$=2$^{\circ}$). 
A small angle oscillatory shear characterization of the samples was carried out at frequencies between 0.1 and 10 Hz. These measurement were taken at low strains ($\gamma=$1\%), after checking that this was within the linear viscoelastic range. At least three repetitions were performed for each sample.

Additionally the IDDSI flow test \cite{Cichero2016} was run in triplicates at the temperature of 22$^{\circ}$C. This consists in measuring the volume of fluid discharged through the orifice of standard eccentric luer slip tip 10 mL syringe (BD, Franklin Lakes, NJ, USA) in a fixed amount of time (10 s). Depending on the residual liquid hold-up from the initial 10 mL filling level, liquid samples are categorized in four levels of increased thickness. The liquid is classified Level 0 if the syringe is completely empty in 10 s, Level 1 if the remaining liquid volume is between 1 and 4 mL, Level 2 if between 4 and 8 mL and Level 3 if more than 8 mL are left. Thicker liquids than Level 3 do not flow at all through the orifice of the syringe in the prescribed amount of time and are better characterized with the other methods proposed by IDDSI \cite{Cichero2016}.

The extensional properties of the liquids were measured by capillary break-up rheometry using a HAAKE CaBER 1 (Thermo Electron, Karlsruhe, Germany). Five repetitions were perfomed at the temperature of 22$^{\circ}$C.
The instrument measures the thinning of the midpoint diameter of the capillary bridge generated by the rapid separation of two circular plates that axially constrain the liquid sample. The initial separation between the two plates of diameter $D_0$=6 mm was set at $h_0$=3 mm. This gives an initial aspect ratio ${h_0}/{D_0}$=0.5, value that limits the initial curvature of the liquid column resulting from the combined effect of gravity ($g$) and surface tension ($\sigma$) \cite{Yao1998, Rodd2005}. The liquid sample was injected between the plates using a 1 mL syringe. The experiment was triggered 60 s after loading the sample, to avoid shear preconditioning effects \cite{Bhardwaj2007}.

An axial displacement (to $h_f$=13 mm) was imposed in 50 ms to drive the filament thinning. As the capillary thread thins, its midpoint diameter is measured with a laser micrometer with a beam thickness of 1 mm and a resolution of 20 $\mu$m. High-speed videos of the experiment were taken at 1000 frames per second to record the shape evolution of the capillary thread using a Phantom V1612 high-speed camera (Vision Research, Wayne, NJ) equipped with a telecentric TC16M012 f/18 lens (Opto Engineering, Mantova, Italy). The resulting image resolution for the optical configuration installed was approx. 9.6 $\mu$m/pixel.

The cylindrical elements of fluids at the axial mid-plane plate are subject to a strain expressed by Eq.~\ref{Eq_NEW}.

\begin{equation}
\varepsilon=-2\ln{\frac{D_{mid}}{D_0}}
\label{Eq_eps}
\end{equation}

Where $D_{mid}$ is the instantaneous filament midpoint diameter. 

The filament necking, driven by capillarity, is opposed by viscous, inertial and elastic terms. The relative importance of these contributions depends on the rheological and physical properties of the sample, other than the characteristic size of the filament \cite{Sur2018}. The relative importance between viscosity and surface tension forces is quantified through the Ohnesorge number,

 $Oh=\eta/ \left (\rho\, \sigma\, 0.5\,D_{mid} \right )$, 
 
 where $\eta$ is the shear viscosity, $\rho$ the liquid density and $\sigma$ the surface tension \cite{Mckinley2005,Mathues2015}.

A theory describing the time evolution of slender fluid filaments, in the limit of vanishing inertial terms was presented by Renardy \cite{Renardy1995} and analytical solutions have been presented for simple constitutive models.
In case of Newtonian fluids, the visco-capillary force balance predicts a linear rate of decay of the midpoint diameter, function of the capillary velocity, defined as the ratio of surface tension and shear viscosity (Eq.~\ref{Eq_NEW}). Conversely, for an upper convected Maxwell model (UCM) the elasto-capillary force balance predicts an exponential diameter decay in time with a characteristic time scale $\lambda_c$ (Eq.~\ref{Eq_UCM}). This constant was linked to the longest relaxation mode in shear for aqueous polyethylene-oxide solutions \cite{Rodd2005}.

\begin{equation}
D_{mid} \sim \frac{\sigma}{\eta}\left ( t_c-t \right )
\label{Eq_NEW}
\end{equation}

\begin{equation}
D_{mid} \sim \sqrt[3]{\frac{D_0^2}{\sigma}} \exp\left ( -{\frac{t}{3 \lambda_c}}
\right )
\label{Eq_UCM}
\end{equation}

The instantaneous strain rate for a cylindrical element of fluid is given by Eq.~\ref{Eq_EPS} and is a constant, equal to $2/(3\,\lambda_c)$, for ideal elastic (i.e. UCM) liquids \cite{Mckinley2005, Mathues2015}.

\begin{equation}
\dot{\varepsilon}=-\frac{2}{D_{mid}}\frac{dD_{mid}}{dt}
\label{Eq_EPS}
\end{equation}

The apparent extensional viscosity of the liquid  can be expressed by
Eq.~\ref{Eq_ETA} \cite{Mckinley2005, Mathues2015}.

\begin{equation}
\eta_{app}^{ext}=\frac{2 \sigma/D_{mid}}{\dot{\varepsilon}}
\label{Eq_ETA}
\end{equation}

In case of non/slender liquid filaments, the analysis of the self-similar solution proposed by Papageorgiou \cite{Papageorgiou1995}, and summarized by Mckinley \cite{Mckinley2005}, leads to the introduction of a corrective factor in Eq.~\ref{Eq_ETA}. This prefactor, that accounts for the axial variations in the filament shape during visco-capillary thinning, is approximately equal to 0.43 for Newtonian liquids \cite{Mathues2015}. 


Surface tension was measured using a tensiometer Kruss ST10 (KRÜSS GmbH, Hamburg, Germany) using a Wilhelmy plate. Five repetitions were taken per each sample.

\subsubsection{The \textit{in vitro} model of swallowing} 
An \textit{in vitro} experiment was used to assess the effect of the bolus rheology on the characteristic oral transit time and on the bolus shape, in conditions relevant to oral swallowing. The experimental setup simplifies the \textit{in vivo} peristaltic motion induced by the tongue during the oral phase of swallowing in a quasi-two-dimensional geometry. A comprehensive description of the experimental setup was given by Mowlavi \textit{et al.} \cite{Mowlavi2016}. A 6 mL bolus volume was used for all the \textit{in vitro} experiments, equivalent to 10mL  \textit{in vivo} \cite{Mowlavi2016}. \textcolor{black}{The role of the salivary lubrication has not been considered in this study and this may limit the relevance of the results to hyposalivation.}

The \textit{in vitro} swallowing experiment was recorded using a high-speed camera (model ac1920-155 $\mu$m, Basler, Isny im Allgäu, Germany) to measure the instantaneous position of the bolus leading edge and its length. Two key events were identified: bolus front out (FO-front out) and bolus tail out (TO-tail out). FO represents the time at which the front of the bolus exits the plastic membrane, which plays the role of the oral cavity in the \textit{in vitro} model. Conversely, TO is the time when the roller impacts the stopper, the tail of the bolus leaves the oral cavity and the peristaltic action stops. 
The evolution of the bolus length along the trajectory and at the outlet of the plastic membrane were calculated from the relative position of bolus front and roller position.
The change in the bolus length was used to compute its elongation during the ejection from the plastic membrane ($ {\varepsilon}_{Bolus}$). The ratio between the bolus elongation during the ejection and the time required for completing the ejection was used to quantify the ejection strain rate ($\dot{\varepsilon}_{Bolus}$).
The injected bolus mass, the mass of bolus ejected after every \textit{in vitro} swallow, and the mass of post-swallow \textcolor{black}{oral} residues left inside the plastic membrane were also recorded for each experiment. This parameter was found to correlate well with sensory data considering the in mouth grittiness and swallow after-feel of placebo formulations \cite{MarconatiUCL}.

Theoretical predictions of the swallowing dynamics were derived from the Lagrangian equation of motion of the mechanical system considering the viscous dissipation of the bolus as a modified Poiseuille drag flow \cite{Hayoun2015a,Mowlavi2016}. Although idealized, this approach offered remarkably good predictions and helped understanding the effect of a shear thinning rheology on bolus oral transit time.  The theory presented in Hayoun \textit{et al.} \cite{Hayoun2015a} was extended in this study to consider shear thinning rheological behaviours for fluids whose steady shear viscosity is well fitted either by Power-law or Ellis models, expressed by Eq.~\ref{EqPL} and Eq.~\ref{Eq(3)} respectively.

\begin{equation}
\eta_a=~K~\dot{\gamma}^{n-1}
\label{EqPL}
\end{equation}

\begin{equation}
\eta=\frac{\eta_0}{1+\left | \frac{\tau}{\tau_0} \right |^{\alpha-1}}
\label{Eq(3)}
\end{equation}

Where: $n$ is the power law index, $K$ the consistency index of the Power-law model, while $\eta_0$, $\alpha$, and $\tau_0$ are the parameters of the Ellis model.  

\begin{figure}
\centering
\includegraphics[width=0.5\textwidth]{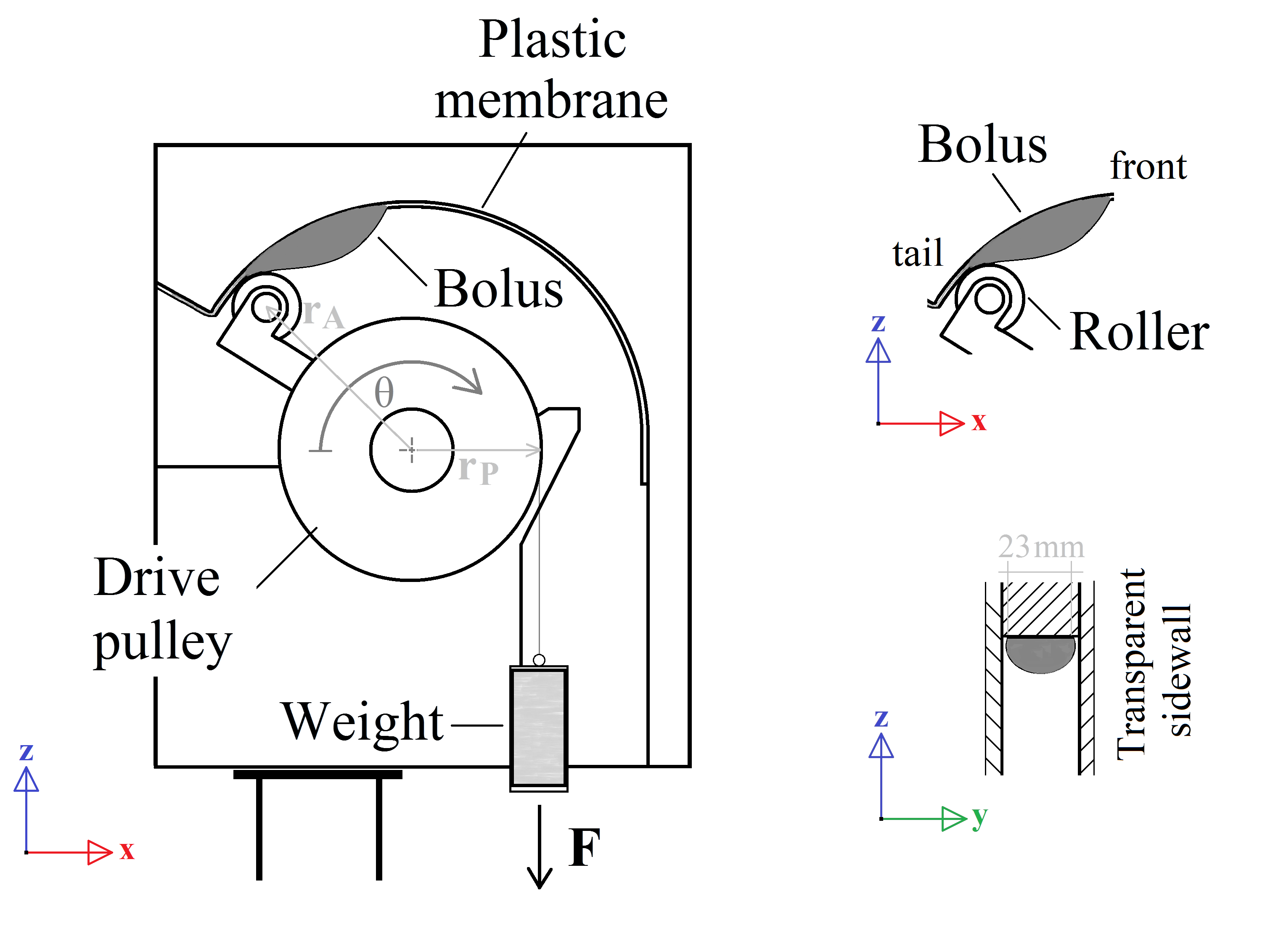}
\caption{\ Schematics of the \text{in vitro} experiment used to replicate the oral phase of swallowing.}
\label{SetupExt}
\end{figure}

\section{Results}

The set of liquids considered in this study was designed to cover a wide range of product consistencies. This was first qualitatively evaluated through the IDDSI flow test, as reported in Table ~\ref{IDDSI}. The results of the flow test indicate how the discharged volume of liquid decreases with increasing polymer concentration. In particular, concentrations of PEO above 3\% w/w led to 
a noticeable amount of residual liquid hold-up in the syringe. Moreover, hardly any outflow was measured in the 10 s test-time for concentrations of PEO above 4\% w/w. 

The 2\% w/w PEO solution, the 1\% w/w cereal extract solution and the TUC solution (1.19 \% w/w) were classified as IDDSI Level 2, the latter in agreement with the level indicated on the package and with the results of other studies \cite{Barbon2018,Dantas2018}. Finally, the 0.3\% w/w cereal extract was characterized as IDDSI Level 1.

\subsection{Shear rheology}
\begin{figure*}[h]
\centering
\includegraphics[width=0.8\textwidth]{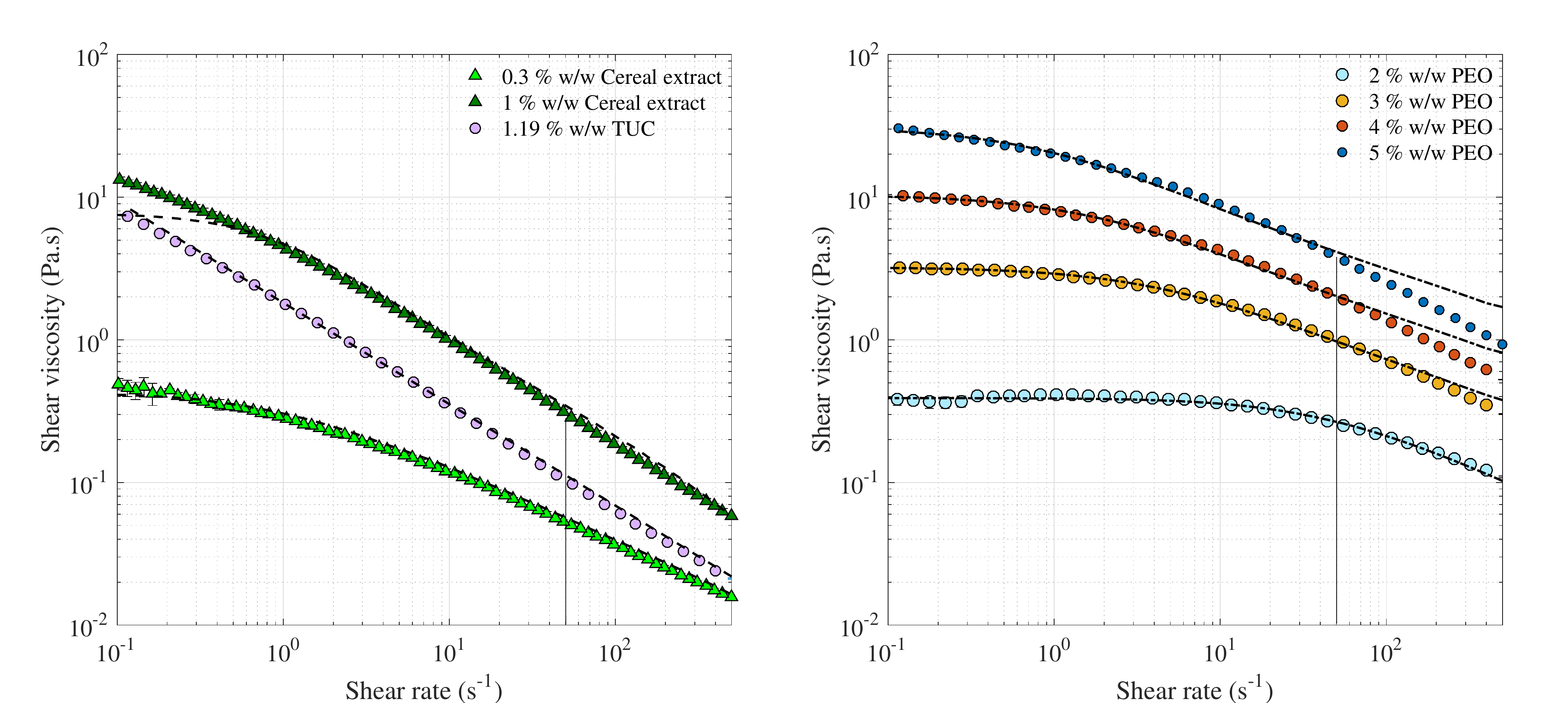}
\caption{\ Steady shear viscosity for aqueous solutions of TUC and a cereal extract (left) and aqueous solutions of 2 to 5 \% w/w PEO (right). The corresponding fitting parameters of are listed in Table~\ref{Fitting}.}
\label{ShearTUC}
\end{figure*}

Rheological measurements in steady shear showed a shear thinning behaviour for all products, although with specific differences. The \textit{Nectar-thick} solution of TUC showed a pronounced shear thinning behaviour well-fitted by a power law model in the range of shear rates considered (Table~\ref{Fitting}, reported in the appendix). In line with previous studies \cite{Ebagninin2009}, the aqueous solutions of PEO are less shear thinning than TUC and highlight a visible a high viscosity plateau at low shear rates. The breadth of this viscosity plateau increases with polymer concentration (Fig.~\ref{ShearTUC} and \ref{ShearTUC}). 
\textcolor{black}{The TUC solution, the 1\% w/w cereal extract solution and the 2\% w/w PEO (all IDDSI Level 2) show a similar shear viscosity at the shear rates of 50 and 300 reciprocal seconds.}

The flow curves of the cereal extracts show an intermediate behavior between TUC and PEO, being slightly more shear thinning than PEO but less than the gum-based thickener (Fig.~\ref{ShearTUC} and \ref{ShearTUC}). The Ellis model (Eq.~\ref{Eq(3)}) fits auite well the flow curves of the PEO and the cereal extract solutions in the range of shear rates considered (Table \ref{Fitting}). The model captures well the low shear rate Newtonian plateau and the onset of the shear thinning region (Fig.~\ref{ShearTUC}), as demonstrated by the R$^2$ values reported in Table \ref{Fitting}.

\begin{table*}
\small
\centering
\caption{\ Results for the IDDSI flow test. Depending on the amount of residual liquid volume left in the syringe from the initial 10 mL filling level, the liquid samples are categorized into four levels of increased thickness: Level 0 if the syringe is completely empty in 10 s, Level 1 if the remaining liquid volume is between 1 and 4 mL, Level 2 if between 4 and 8 mL and Level 3 if more than 8 mL are left. Thicker liquids than Level 3 do not flow at all through the orifice of the syringe in the prescribed amount of time. These measurements were repeated 3 times and the standard deviation is indicated in brackets.}
\label{IDDSI}
\begin{tabular*}{\textwidth}{@{\extracolsep{\fill}}ccccccc}
\hline
Liquid sample & Remaining volume & IDDSI level \\
\hline
1.19 \% w/w TUC & 4.9 (0.3) & Level 2 \\
0.3 \% w/w cereal extract & 2.9 (0.3) & Level 1 \\
1 \% w/w cereal extract & 5.3 (0.2) & Level 2 \\
2\% w/w PEO & 7.0 (0.2) & Level 2 \\
3\% w/w PEO & 9.5 (0.2) & Level 3 \\
4\% w/w PEO & 9.9 (0.1) & Level 3 \\
5\% w/w PEO & 10.0 (0.1) & Level 4 \\
\hline
\end{tabular*}
\end{table*}

\begin{figure*}
\centering
\includegraphics[width=0.8\textwidth]{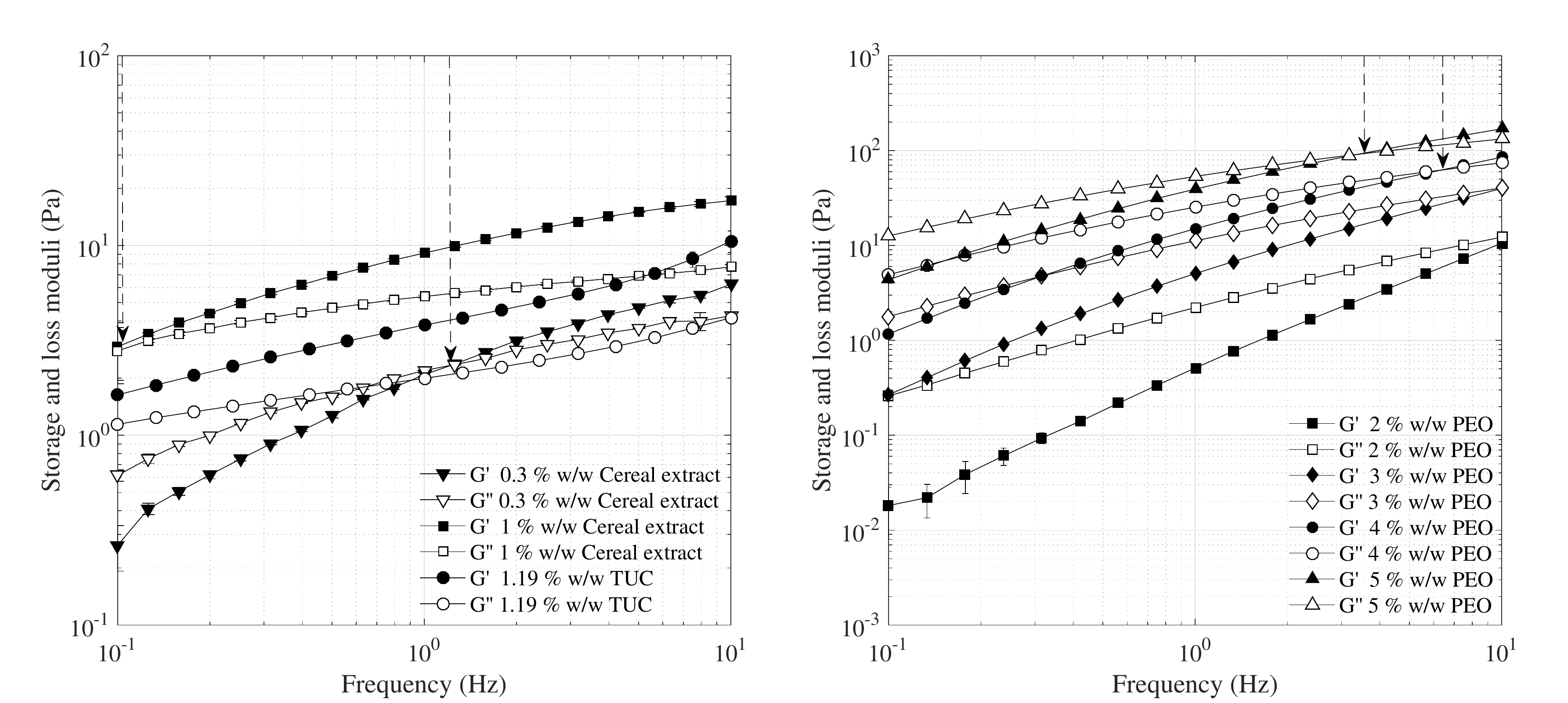}
\caption{\ Oscillatory shear tests at 1\% strain for TUC \textit{Nectar-thick} and aqueous solutions of a cereal extract (left) and aqueous solutions of 2 to 5 \% w/w PEO (right). Storage (G') and Loss (G'') moduli are illustrated with filled and open markers respectively. Whenever observable, crossover frequencies are indicated with an arrow.}
\label{SAOSTUC2}
\end{figure*}

The result of the small amplitude frequency sweeps within the linear viscoelastic regime is illustrated in Fig.~\ref{SAOSTUC2}. The TUC solution shows a dominant storage modulus (G') over the loss modulus (G'') in the range of frequency considered and the shear moduli of TUC do not exhibit a strong frequency dependence: a behavior that is consistent with previous results from Mackley \textit{et al.} \cite{Mackley2013}. Conversely, in case of PEO and cereal extract, G'' dominates the first part of the mechanical spectra with a crossover occurring at higher frequencies. 

\subsection{Extensional rheology}
\begin{figure}
\centering
\includegraphics[width=0.45\textwidth]{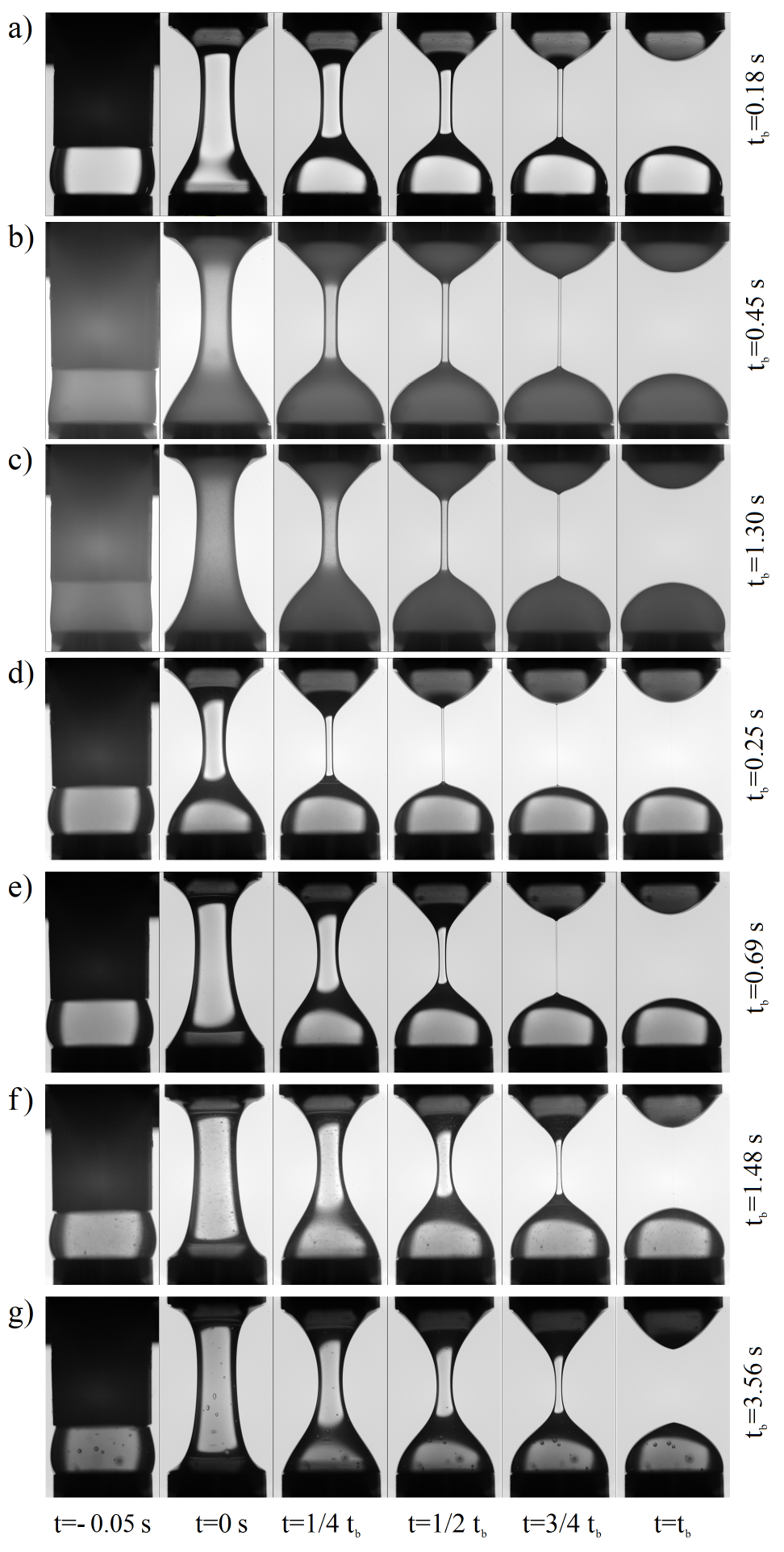}
\caption{\ Capillary thinning for TUC and cereal extract solutions: row a) 1.19\% w/w TUC (\textit{Nectar-thick}), b) 0.3\% w/w cereal extract, c) 1\% w/w cereal extract, d) 2\% w/w PEO and e) 3\% w/w PEO f) 4\% w/w PEO g) 5\% w/w PEO. Times are normalized with respect to the capillary break-up time, reported on the right.}
\label{Capillary}
\end{figure}

\begin{figure*}
\centering
\includegraphics[width=0.8\textwidth]{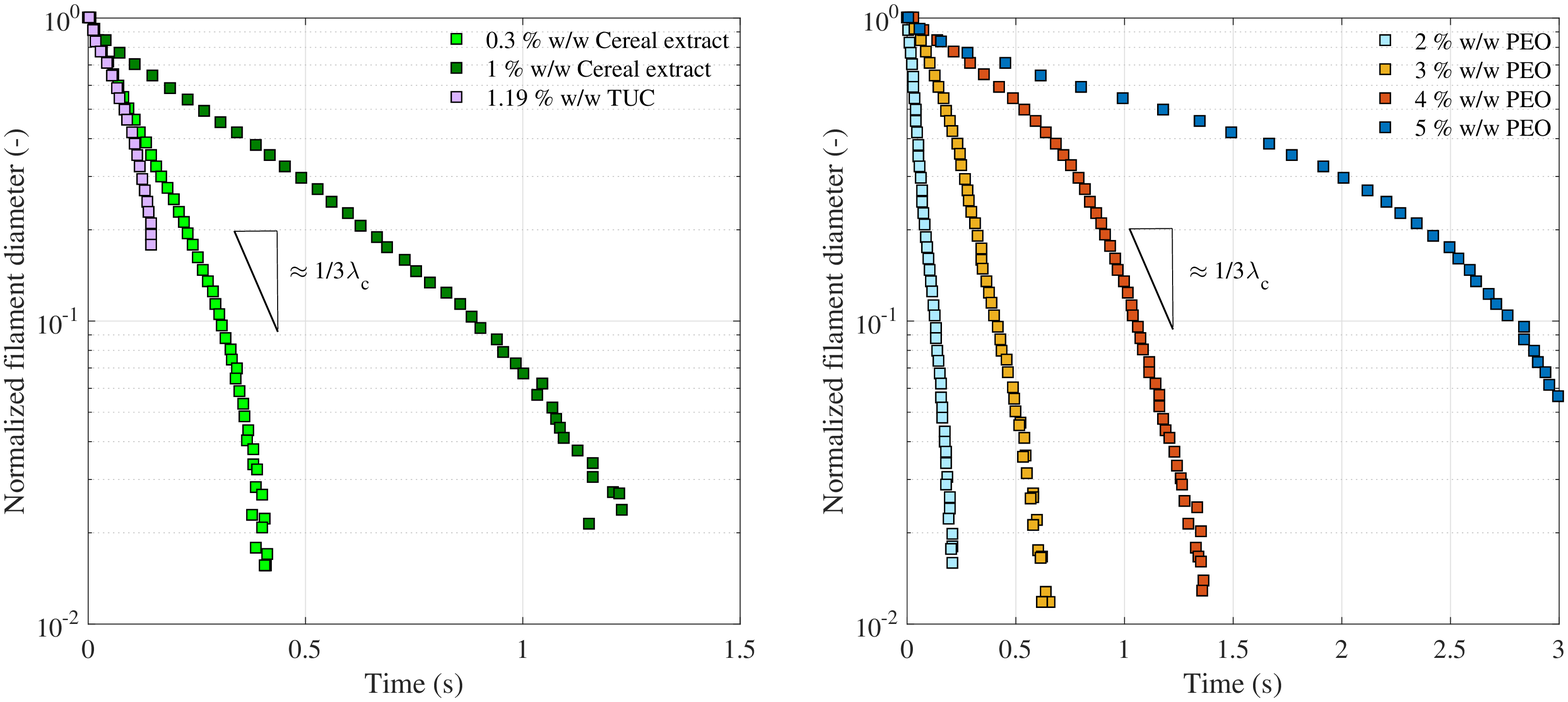}
\caption{\ Profiles of the capillary thinning for aqueous solutions of TUC and a cereal extract (left) and aqueous solutions of 2 to 5 \% w/w PEO (right).}
\label{CaberTUC}
\end{figure*}

The extensional properties of the samples were studied by capillary break-up. Selected images taken during the transient thinning of the filaments until break-up are presented in Fig.~\ref{Capillary} while Fig.~\ref{CaberTUC} illustrates the evolution with time of the filament diameter, normalised by the initial diameter. Five repetitions were performed for each sample. The filament break-up times span a range between 0.1 and 3.5 s (Table \ref{Break-up}) \textcolor{black}{and always increase with increasing concentration. The breakage times of the 1\% w/w cereal extract solution is significantly higher than that of the two other IDDSI Level 2 solutions. Interestingly, even the breakage times of the (IDDSI Level 1) 0.3\% w/w cereal extract exceeds that of the other two Level 2 solutions.}

The video recordings of the capillary break-up allow to appreciate the different regimes of filament thinning. This dynamic process is driven by capillarity and resisted by viscous, elastic and inertial forces. For Newtonian fluids the viscous response is highlighted by a hour-glass filament shape, rapidly evolving with time. In this case, the filament diameter decreases linearly in time \cite{Anna2001a}. Conversely, elastic fluids thin with an approximately cylindrical capillary, whose radius exponentially decreases in time \cite{Anna2001a}. This last behavior is well-matched by the cereal extract solutions (Fig.~\ref{CaberTUC}). 

In this case a single relaxation time is adequate to describe the capillary thinning dynamics, that is predominantly confined to the elasto-capillary regime. Such relaxation time is proportional to the inverse of the gradient and is of the order of 0.05 to 0.15 s (Table \ref{Break-up}).

A similar single-mode elastic capillary thinning is also observed for the aqueous solutions of 2 and 3\% w/w PEO. However, at higher PEO concentrations, the viscous drainage gives a more significant contribution to the first portion of the filament thinning dynamics. This effect can qualitatively be appreciated comparing the filament shape in Fig.~\ref{Capillary}b and d. In particular, results for the 5\% PEO show a distinctive long time decay of the filament axial curvature (Fig.~\ref{Capillary}).

In case of TUC, the dynamics of filament break-up is complicated by the combined effect of the elastic contribution and the shear thinning rheology. In the viscous dominated regime, this results in an acceleration of filament break-up as opposed to less shear thinning liquids. In this case, a spectrum of relaxation times, rather than a single one would be more appropriate for a more comprehensive description of the extensional flow \cite{Mckinley2005,Anna2001a}. 

\begin{figure*}
\centering
\includegraphics[width=0.8\textwidth]{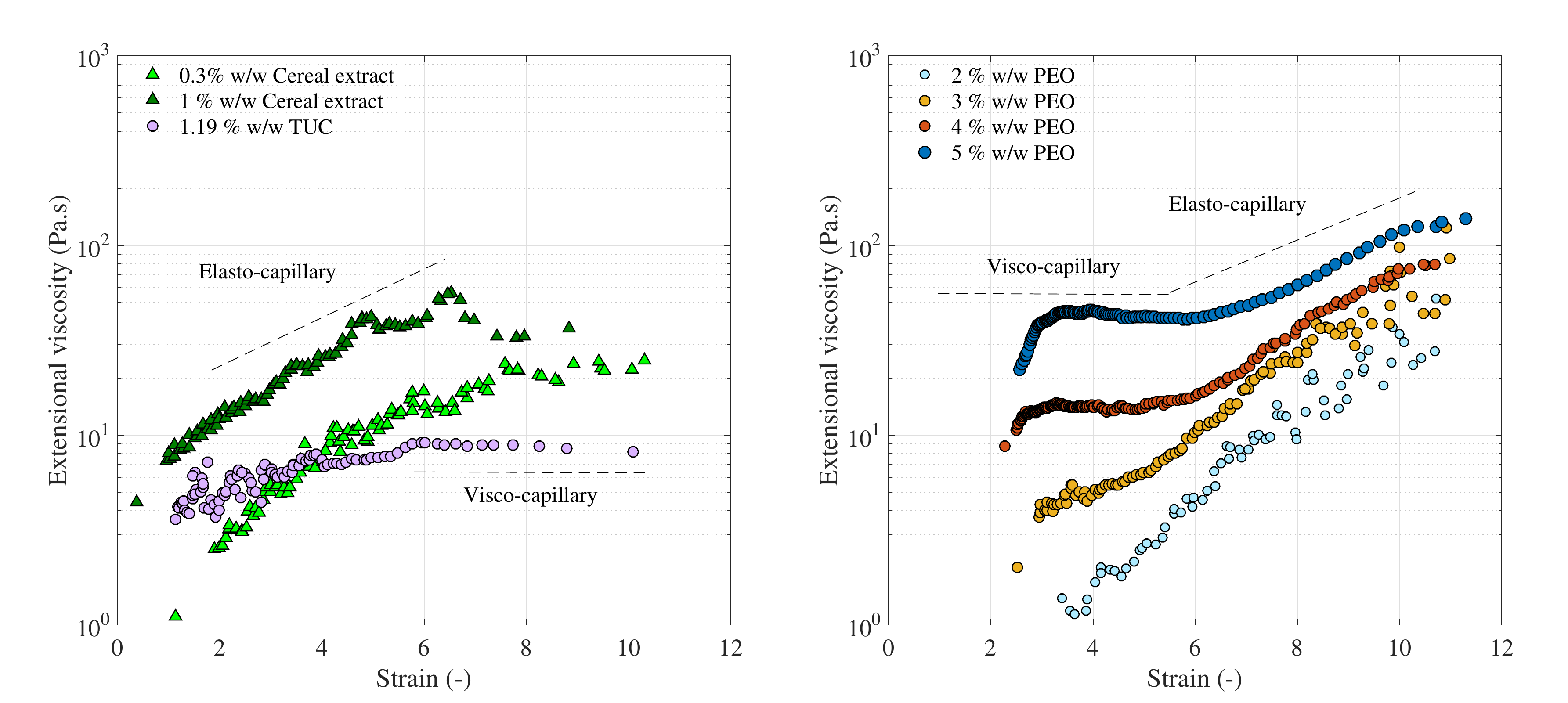}
\caption{\ Values of apparent extensional viscosity as a function strain for aqueous solutions of TUC and a cereal extract (left) and aqueous solutions of 2 to 5 \% w/w PEO (right).}
\label{StrainTUC}
\end{figure*}

Inertial forces affect the dynamics of capillary thinning only for low viscosity fluids, for which the Ohnesorge number falls into the inertio-visco-capillary regime ($Oh<0.2$) \cite{Mckinley2005,Mathues2015}. This condition was however not observed for any of the liquids tested.

The apparent transient extensional viscosity of the fluids tested was calculated using Eq.~\ref{Eq_ETA}. The measured value of surface tension are rather similar for all the fluids and show only a weak dependence on concentration (Table \ref{Break-up}).
The calculated values of apparent extensional viscosity are reported in Fig.~\ref{StrainTUC} as a function of strain. TUC shows a different behaviour with respect to the other thin (viscoelastic) solutions. The weak strain dependence of the TUC solution is compatible with a dominant visco-capillary thinning dynamics that is also highlighted by the quicker breakage of the filament, compared both to the PEO and the cereal extract solutions tested. The strain hardening behaviour of PEO and the cereal extract solutions confirms the more important contribution of elasticity during the transitory filament break-up, a phenomenon that is typical of elastic fluids (Fig.~\ref{StrainTUC}). Results for TUC and PEO are close to the values reported in literature \cite{Hadde2019, Arnolds2010} and the apparent extensional viscosity of the cereal extracts are within the range of 2 to 4\% PEO (Fig.~\ref{StrainTUC}).
The increase in extensional viscosity with strain observed for the cereal extracts and the PEO solutions follows a similar exponential growth in the elasto-capillary regime.
Furthermore, Fig.~\ref{StrainTUC} does not highlight a strong relationship between the maximum attainable strain and the polymer concentration.
This is in contrast with Arnolds \textit{et al.} that reported a noticeable increase in the maximum attainable strain with increasing PEO concentration due to a bead-on-a-string instability in the final phases of filament break-up \cite{Arnolds2010}.

The exponential diameter decay of the cereal extract solutions occurs at approximately constant strain rates within the elasto-capillary regime of the filament thinning. 
The theoretical strain rate of the extensional flow can be computed as $2/(3\, \lambda_c)$ \cite{Anna2001a,Mathues2015}.
In case of PEO, a good agreement is also found between the measured relaxation times for the extensional flow and the scaling rule with polymer concentration $c$, proposed by Arnolds \textit{et al.} \cite{Arnolds2010} and expressed by $\lambda_c \sim c^{4/3 \pm1/3}$.

The transient Trouton ratio ($Tr$), calculated as the ratio of extensional to the zero shear viscosity \cite{Anna2001a} (Figure~\ref{TrPEO} and Table~\ref{Fitting}), enables to quantify the visco-elasticity of the liquid solutions tested. The cereal extract solutions show higher Trouton ratios than TUC for comparable IDDSI concistency, thus confirming the significantly higher elasticity. 

A decrease in $Tr$ is observed increasing the concentration of PEO and this is due to the sharp increase of zero shear viscosity with concentration that dominates over the increase in extensional viscosity \cite{Arnolds2010}. For thick PEO solutions, such as 4 and 5\% PEO, $Tr$ is close to the theoretical limit of 3, expected for Newtonian fluids. The same observation applies to the cereal extract solutions that exhibit also some decrease in $Tr$ with increasing polymer concentration.


\begin{table*}
\centering
\small
\caption{\ Capillary break-up time ($t_{b}$), extensional relaxation time $\lambda_{c}$ and surface tension ($\sigma$) of the liquids tested. Standard deviation in brackets.}
\label{Break-up}
\begin{tabular*}{\textwidth}{@{\extracolsep{\fill}}cccc}
\hline
Liquid sample & $t_{b}$ (s) & $\lambda_{c} $ (s) & $\sigma$ (mN/m) \\
\hline
1.19 \% w/w TUC & 0.18 (0.02) & 0.02 (0.00) & 61.5 (0.9) \\
0.3 \% w/w cereal extract & 0.45 (0.02) & 0.05 (0.01) & 53.3 (0.7) \\
1 \% w/w cereal extract & 1.30 (0.05) & 0.14 (0.01) & 57.0 (0.9) \\
2\% w/w PEO & 0.25 (0.01) & 0.02 (0.00) & 62.2 (0.7) \\
3\% w/w PEO & 0.69 (0.04) & 0.05 (0.01) & 53.7 (1.1) \\
4\% w/w PEO & 1.48 (0.03) & 0.06 (0.02) & 55.4 (0.9) \\
5\% w/w PEO & 3.56 (0.26) & 0.11 (0.04) & 56.3 (1.7) \\
\hline 
\end{tabular*}
\end{table*}

\subsection{\textit{In vitro} swallowing tests}
The \textit{in vitro} swallowing tests gave quantitative information on the swallow-ability of the samples in terms of oral transit velocity and mass of \textcolor{black}{oral} residues.

\begin{figure*}
\centering
\includegraphics[width=0.95\textwidth]{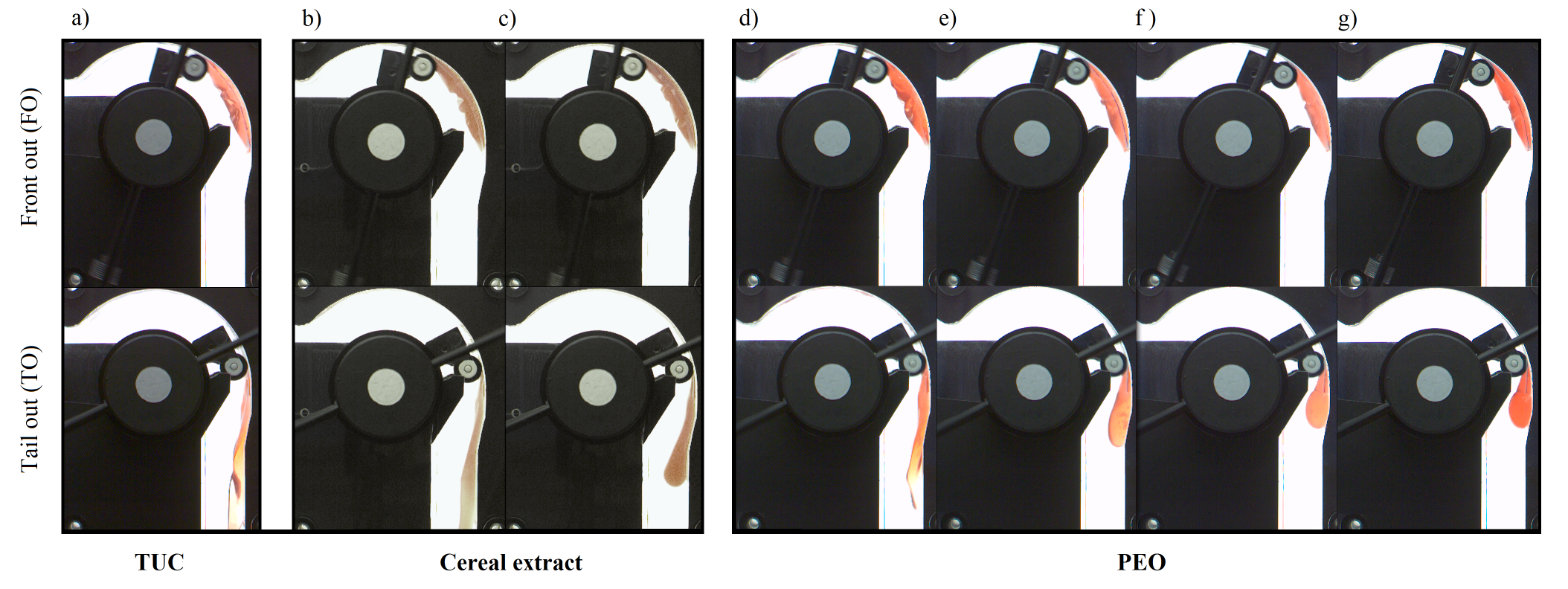}
\caption{\ Relevant pictures taken during the \textit{in vitro} experiments. From left to right the instant of bolus front out (top row) and tail out (bottom row) for a) 1.19\% w/w TUC (\textit{Nectar-thick}), b) 0.3 \% w/w cereal extract, c) 1 \% w/w cereal extract and d)-g) aqueous solutions of 2 to 5\% w/w PEO. \textcolor{black}{Solutions a), c) and d) were all characterised as IDDSI Level 2 and showed similar shear viscosity at 50 and 300 s-1, but the solution c) showed significantly stronger extensional properties. These snapshots show that the \textit{in vitro} bolus of solution c) is more compact when leaving the oral cavity}.}
\label{FOTUC}
\end{figure*}

\begin{figure}
\centering
\includegraphics[width=0.45\textwidth]{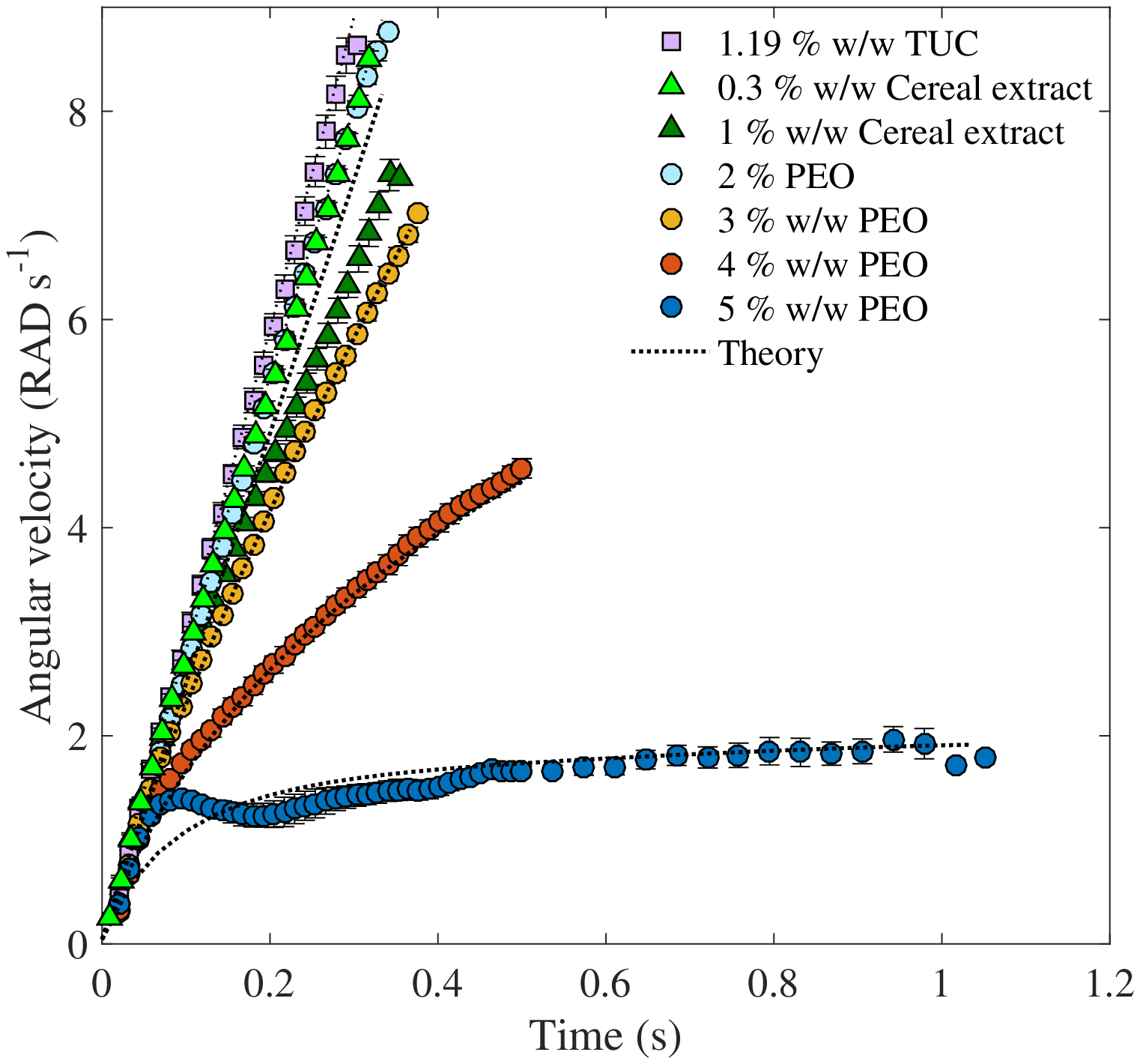}
\caption{\ Evolution of the angular velocity of the tail of the bolus and of the roller in the \textit{in vitro} experiments. The TUC solution, the 0.3 \% w/w cereal extract and 2 \% w/w PEO) show a dynamics governed by the bolus and tongue inertia.}
\label{RollerPEO}
\end{figure}

Pictures taken when the head of the bolus (FO) and the tail of the bolus (TO) leave the \textit{in vitro} oral cavity are reported in Fig.~\ref{FOTUC} for the different liquids considered in this study.
\textcolor{black}{The bolus velocity during the oral transit was calculated from the instantaneous positions of the roller, as described in more detail in previous publications \cite{Hayoun2015a}. A close examination of the bolus velocity profiles provides an important insight on the relative importance of the viscous dissipation, controlled by the liquid rheology, vs. the bolus and tongue inertia  (Fig.~\ref{RollerPEO})}. Whenever the viscous dissipation in the bolus becomes negligible, the roller and bolus mean velocity increase linearly with time. The angular acceleration is then constant and only dependent upon the inertia of the system, mimicking the tongue inertia. The dynamics is also almost insensitive to the bolus mass and viscosity \cite{Hayoun2015a}. Conversely, whenever viscous dissipation is sufficiently high, the system evolves towards constant values of angular velocity after an initial, short, inertial phase. In the case of shear thinning liquids, the transition from the inertial to the viscous regime depends on the shear rates generated during the bolus flow. This effect is illustrated in Fig.~\ref{RollerPEO} where the experimental velocity profiles are plotted until the moment in which the bolus front exits the \textit{in vitro} oral cavity (FO). Experiments with the thicker solution of PEO result in values of angular velocity approaching a constant, while thin and highly shear thinning solutions always evolve within the inertial regime. The behaviour of both fluids is remarkably well captured by the theoretical model that only considers shear viscosity, neglecting elasticity.

In between the extremes situation of 5\% PEO and TUC solution, the set of experiments with 1 to 4\% PEO gives significant insights on the importance of elastic properties during swallowing. Solutions for the 3\% and 4\% PEO show a two folds difference in apparent shear viscosity measured at 50 reciprocal seconds. Whilst the former evolve always within the inertial regime, the latter shows a strong effect of the viscous dissipation on the bolus dynamics. A threshold shear rate to compare the effect of viscosity cannot however be univocally defined as the maximum shear rates generated during swallowing are a strong function of the shear rheology of the bolus. 

The mass of post-swallow \textcolor{black}{oral} residues left in the plastic membrane was measured after each experimental run. \textcolor{black}{The average residue left by the 1.19 \% w/w TUC solution was used as a reference to normalize the results obtained with the other fluids. All normalized results are reported in Table \ref{Transit}.
The residues increase with concentration. This effect is strong for the PEO solutions and weaker for the cereal extract solutions.} Among the liquids tested, the low concentration cereal extract sample generated the lowest residues. This observation suggests a possible advantage of thin elastic liquids in reducing the quantity of oral residues after swallowing.

The \textit{in vitro} model also allows measuring the evolution with time of the bolus shape. Both the transit time and the shape of the bolus depend strongly on the liquid considered. 
The loss of confinement occurring at the exit of the plastic membrane represents the transition from the oral to the pharyngeal phase of swallowing. During this transition, it is instructive to analyse the bolus length.
For both cereal extracts and for the TUC solution the bolus ejection from the \textit{in vitro} oral cavity is followed by an increase in bolus length (Fig.~\ref{FOTUC} and Table \ref{Elongation}). However, as the concentration of polymer is increased, images reveal a more compact and shorter bolus. 
For instance, increasing the concentration of cereal extract results in a three-fold decrease in the ejection strain when the bolus leaves the \textit{in vitro} palate (Table \ref{Elongation}). It can also be observed that the surface of the cereal extract bolii look smoother than the surface of the TUC boli.

\begin{figure}
\centering
\includegraphics[width=0.45\textwidth]{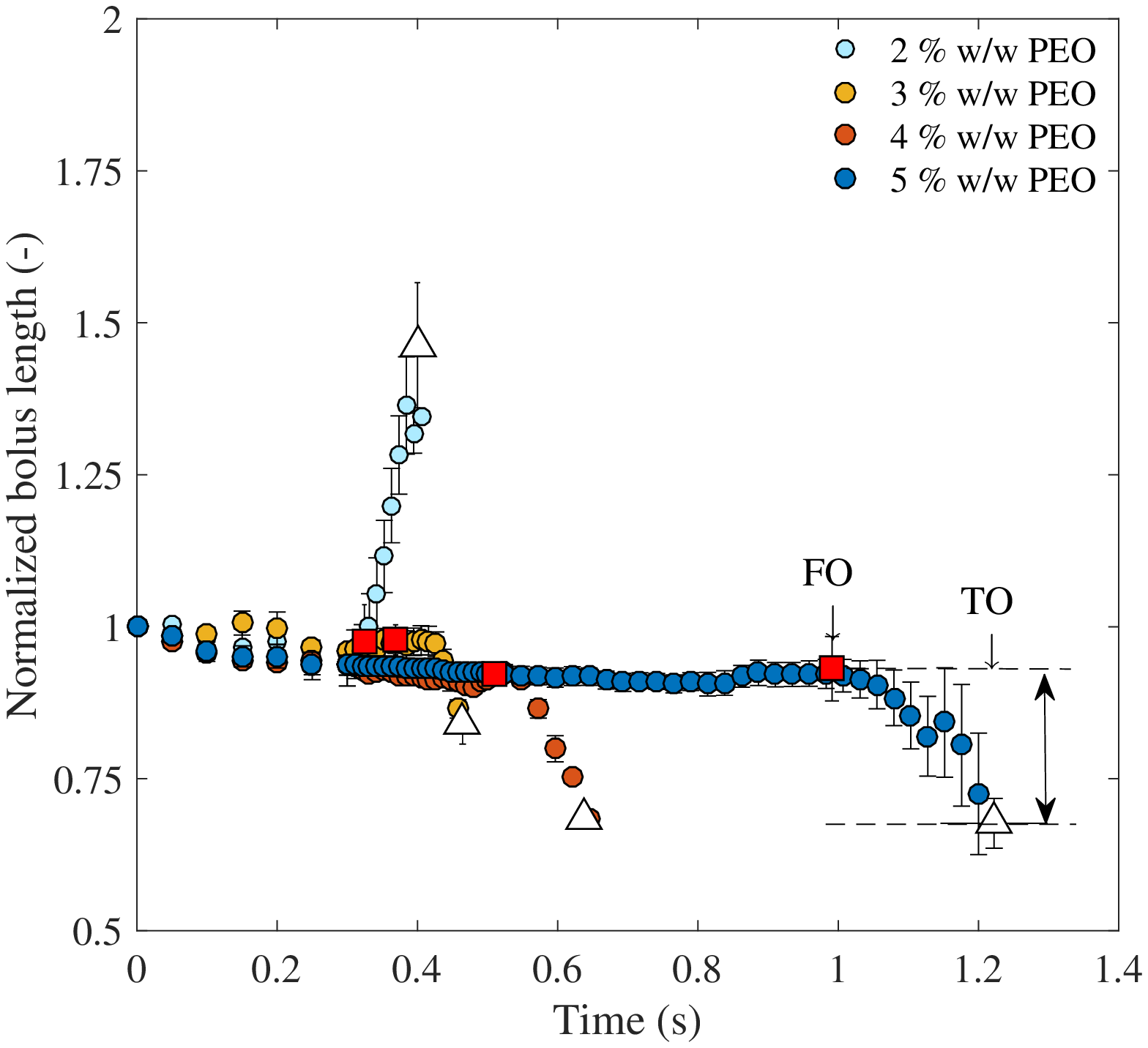}
\caption{\ Normalized bolus length for aqueous PEO solutions. Average values and maximum deviation from three repetitions.}
\label{LengthPEO}
\end{figure}

\begin{table*}[htp]
\small
\centering
\caption{\ Post-swallow residues, characteristic transit oral transit time (FO), time for bolus ejection (TO-FO) and angular velocity of the roller at FO from \textit{in vitro} tests. The amount of post-swallow residues is normalized with respect to TUC \textit{Nectar-thick}. Average values over three repetitions. Standard deviation in brackets.}
\label{Transit}
\begin{tabular*}{\textwidth}{@{\extracolsep{\fill}}ccccc}
\hline
Liquid sample & Normalized residues (-) & FO (s) & \multicolumn{1}{l}{TO-FO (s)} & $\dot{\theta}$ at FO (RAD/s) \\ \hline
1.19 \% w/w TUC & 1.00 (0.07) & 0.30 (0.01) & 0.09 (0.01) & 8.96 (0.24) \\
0.3 \% w/w cereal extract & 0.78 (0.13) & 0.32 (0.01) & 0.09 (0.01) & 8.57 (0.14) \\
1 \% w/w cereal extract & 1.11 (0.19) & 0.35 (0.01) & 0.09 (0.02) & 7.61 (0.08) \\ 
2\% w/w PEO & 1.59 (0.16) & 0.33 (0.01) & 0.08 (0.02) & 8.92 (0.11) \\
3\% w/w PEO & 1.84 (0.09) & 0.38 (0.01) & 0.09 (0.02) & 7.15 (0.15) \\
4\% w/w PEO & 2.34 (0.17) & 0.52 (0.01) & 0.12 (0.02) & 4.76 (0.13) \\
5\% w/w PEO & 2.67 (0.23) & 1.01 (0.05) & 0.23 (0.03) & 2.10 (0.20) \\ 
\hline
\end{tabular*}
\end{table*}

\begin{table*}[htp]
\centering
\small
\caption{\ Characterisation of the bolus ejection from the oral cavity: bolus elongation ($\epsilon_{Bolus}$), strain rate ($\dot{\epsilon_{Bolus}}$) and non-dimensional Deborah ($De$), Reynolds ($Re$), and Elasticity ($E$) numbers were computed during the transition from front out (FO) to tail out (TO). Average values over three repetitions. Standard deviation in brackets.}
\label{Elongation}
\begin{tabular*}{\textwidth}{@{\extracolsep{\fill}}cccccc}
\hline
Liquid sample & $\epsilon_{Bolus}$ (\%) & $\dot{\epsilon_{Bolus}}$ (s$^{-1}$) & $De$ (-) & $Re$ (-) & $E$ (-) \\ \hline
1.19 \% w/w TUC & 41.18 (14.09) & 4.75 (0.10) & 0.22 & 113 & 2E-3 \\
0.3 \% w/w cereal extract & 64.86 (17.63) & 7.04 (1.64) & 0.57 & 163 & 3E-3 \\
1 \% w/w cereal extract & 18.83 (13.06) & 1.90 (0.79) & 1.59 & 3.12 & 0.51 \\ 
2\% w/w PEO & 49.95 (19.87) & 6.61 (0.40) & 0.25 & 27.1 & 0.01 \\
3\% w/w PEO & -13.92 (5.77) & -1.45 (0.08) & 0.56 & 6.32 & 0.09 \\
4\% w/w PEO & -25.90 (2.70) & -2.01 (0.11) & 0.55 & 1.83 & 0.27 \\
5\% w/w PEO & -27.30 (8.53) & -1.19 (0.15) & 0.57 & 0.27 & 1.92 \\
\hline
\end{tabular*}
\end{table*}

Pictures captured when the bolus front (FO) and tail (TO) are leaving the oral cavity show a noticeable lateral expansion (sometiles called elastic recoil or die swell) in case of concentrated PEO solutions, as shown by the evolution of the bolus length during the interval between front out (FO) and tail out (TO) in Fig.~\ref{FOTUC}.
This is illustrated quantitatively in Fig.~\ref{LengthPEO}, where 
a clear shortening of the bolus is visible
for concentrations above 3 \% PEO. This bolus shortening at high concentration is associated with a slower flow (Fig.~\ref{RollerPEO}) that can be explained by the higher shear viscosity and lower Trouton ratio (Figure~\ref{TrPEO}, in Appendix). 

Comparing the results for fluids with similar IDDSI classification highlights some relevant differences in the bolus shape. The IDDSI Level 2 cereal extract solution resulted in a more compact bolus shape and a smoother surface, compared to both 2\% PEO and 1.19 \% TUC (\textit{Nectar-thick}). \textcolor{black}{The smoother surface may suggest a lower risk of surface instabilities leading to undesired fragmentation and this should be confirmed in a geometry representing realistically the anatomy of the human pharynx.}
The higher elasticity and capillary breakage time of the Level 2 cereal extract solution may explain the more compact bolus upon ejection from the \textit{in vitro} oral cavity. Moving to higher IDDSI Levels makes the comparison between the transit dynamics and bolus elongation of different liquids less straightforward, and the role of the shear rheology is amplified.

Die swell has long been studied in continuous processes such as polymer extrusion and blow moulding and several theoretical and semi-empirical equations have been proposed to model the ratio of extrudate to die diameter (swelling ratio) \cite{Tanner2005,Joseph1987,Delvaux1990}. That depends on the ratio between the first normal stress difference and the shear stress calculated at the wall of the die \cite{Tanner2005}. The Deborah and Weissenberg numbers $De$ and $Wi$ are common dimensionless quantities used in the study of viscoelastic flows \cite{Poole2012}. Although conceptually different, in case of transient flows where one length scale determines the dynamics of the flow, they assume a similar definition, based on the ratio between the relaxation time of the polymer, $\lambda$, and a time scale relevant to the flow \cite{Poole2012}. In the study of continuous polymer extrusion, it was shown that the swelling ratio increases with $Wi$, while the position of the maximum die swell is controlled by the elasticity number $E$, defined as the ratio between $Wi$ and the Reynolds number $Re$ \cite{Joseph1987,Delvaux1990}. 

Despite the difference between a continuous extrusion process and the present system, the theory of die swell still allows interpreting some of the \textit{in vitro} results. 
An approximate expression of $De$ for the \textit{in vitro} model here considered can be calculated with the ratio of the extensional relaxation time $\lambda_c$ and the time required for bolus ejection ($De=\lambda_c / ( t_{TO} - t_{FO})$. Increasing the polymer concentration results in higher values of $\lambda_c$ and longer times required for bolus ejection (Table \ref{Break-up} and \ref{Transit}). The ratio $De$ of these solutions scales with the bolus elongation at ejection. For PEO, results show that the calculated values of $De$ increase more than linearly with polymer concentration (Table \ref{Elongation}). For the same polymer, a threshold $De$ for uprising of die swell can be identified: at higher polymer concentrations ($>$3 \% w/w, $De \approx$0.5) the swollen bolus remains coherent and is slowly dragged downward by gravity.
\textcolor{black}{The Deborah number allows also a comparison of the three IDDSI-Level-2 fluids with similar steady shear rheology (the TUC solution, 2\% PEO and 1 \% cereal extract), which also show a similar velocity when exiting the oral cavity. The Deborah number is significantly higher for the cereal extract and this can explain the more compact bolus.
The Deborah number alone cannot however account for the occurrence of die swell across all materials tested, because a different shear viscosity conditions the bolus dynamics and the ejection velocity.}
Overall, thin, elastic and cohesive liquids give rise to a moderate expansion upon leaving the \textit{in vitro} oral cavity (die swell) that partly counteracts the elongation induced by gravity and result in a smooth and compact bolus shape, characterized by a lower bolus elongation. \textcolor{black}{Recent \textit{in vivo} results \cite{Hadde2019_2} provide a first confirmation of the potential benefit on bolus coherence in the pharynx of fluids with some level of extensional viscosity, but further \textit{in vivo} studies should confirm this effect for fluids with stronger extensional properties.
Liquids with similar steady shear rheology show a similar dynamics when leaving the \textit{in vitro} oral cavity, but a more compact bolus is observed when the relaxation time and the filament breakage time in extension are higher.}
A more compact bolus shape is considered a desirable condition \textit{in vivo} to promote a smoother and more controlled bolus flow through the pharynx. 
However experiments show than an excessive bolus swelling, observed for thick elastic fluids, \textcolor{black}{such as concentrated PEO solutions, is associated to an extremely slow bolus transport, which could potentially compromise the airway protection mechanisms.}

\section{Conclusions}
This study provides new insights on the link between the viscoelastic shear and extensional properties of different liquids and their flow during the oral phase of swallowing using an \textit{in vitro} swallowing experiment. Different viscoelastic liquids were tested and a mathematical model was used to predict the experimental bolus velocity profiles from the liquid rheology. 

Highly shear thinning liquids are transported in the \textit{in vitro} oral cavity with a speed that is independent of the shear rheology. With fluids showing higher viscosity at shear rates around 100 reciprocal seconds, the \textit{in vitro} oral transport is conditioned by the steady shear viscosity, but not by the viscoelastic properties in the range considered in this study. 

However, the elastic and extensional properties of the viscoelastic liquid boli play a significant role during bolus ejection from the \textit{in vitro} oral cavity. Thin elastic boli are more compact. \textcolor{black}{The elongation induced by the gravitational acceleration is limited and their bolus surface appears smoother, which may suggest a lower risk of fragmentation that should be confirmed in a geometry representing realistically the anatomy of the human pharynx. For IDDSI-Level 2 boli, presenting a similar shear viscosity and oral dynamics, a higher extensional relaxation time (and filament breakage time) is associated with a more compact bolus.} Thin elastic liquids also show lower \textcolor{black}{oral \textit{in vitro}} residues compared to the other fluids tested.
An even stronger bolus shortening was observed for thick elastic fluids, but this is associated with an undesired slow ejection, due to their high shear viscosity and lower Trouton ratio. 
These findings suggest the importance of a holistic characterization of the rheology of food thickener solutions, \textcolor{black}{including their extensional properties. The rheological properties} can condition the \textcolor{black}{oral} residues, the oral transit time and the shape of the bolus at the transition between the oral and the pharyngeal phase of swallowing. \textcolor{black}{The role of the salivary lubrication has, however, not been considered in this study and this may limit the relevance of these conclusions to hyposalivation.}
\textit{In vitro} models of swallowing  can support the design of novel and safer thickener formulations, although their ultimate validation should naturally come from further clinical trials \textcolor{black}{considering thin boli with significant extensional properties}.

\balance

\section*{Conflicts of interest}
There are no conflicts to declare.

\section*{Acknowledgements}
This study was funded by Nestl\'e Research.

\bibliography{Collection} 
\bibliographystyle{rsc}

\appendix
\renewcommand\thefigure{\thesection.\arabic{figure}}    
\renewcommand\thetable{\thesection.\arabic{table}}    
\pagebreak
\section{Complementary figures}
\setcounter{figure}{0} 
\begin{figure}[htp]
\centering
\includegraphics[width=0.45\textwidth]{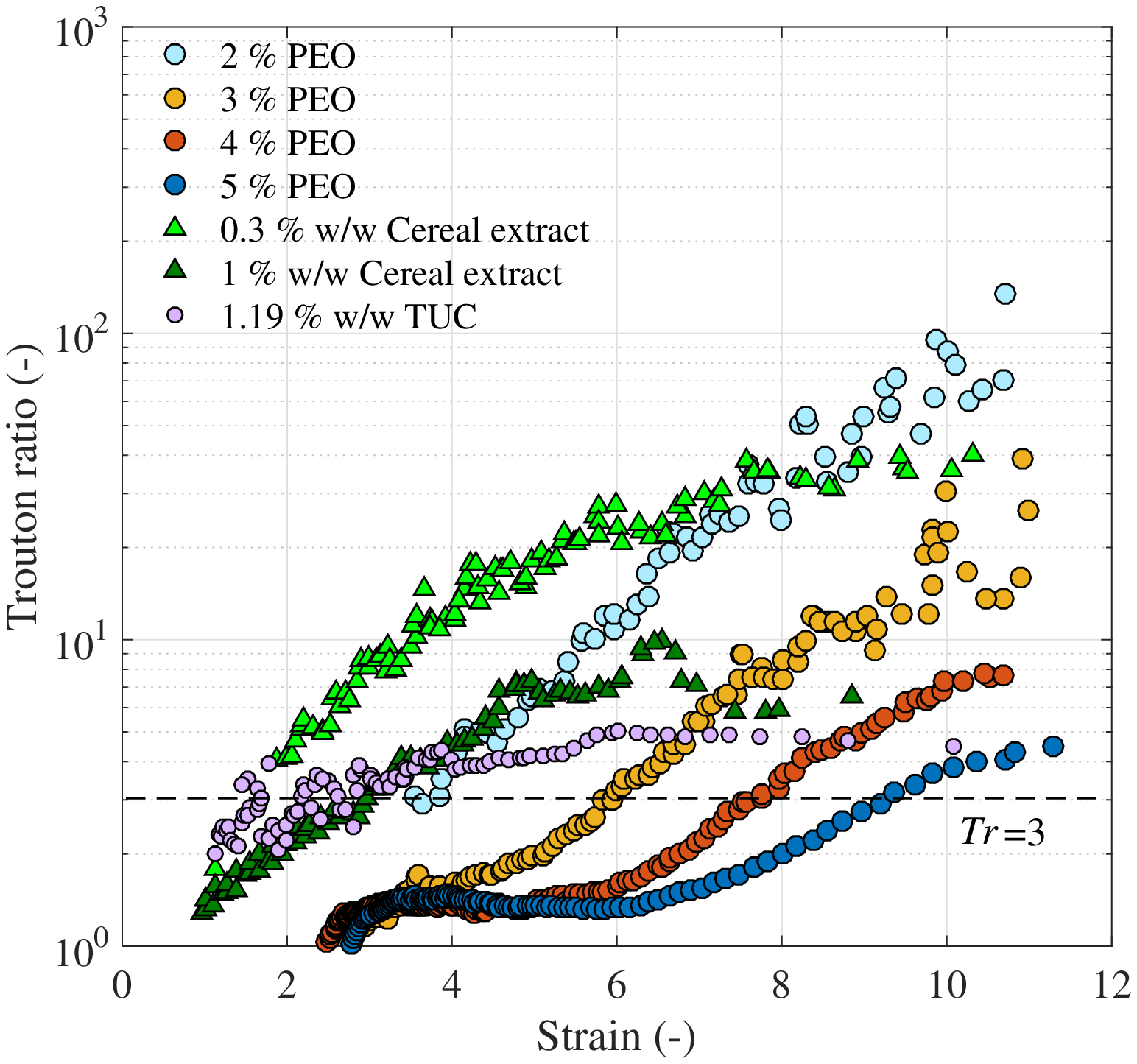}
\caption{\ Ratio of extensional to zero shear viscosity for the liquids considered in this study.}
\label{TrPEO}
\end{figure}

\renewcommand\thefigure{\thesection.\arabic{figure}}    
\renewcommand\thetable{\thesection.\arabic{table}}    
\section{Complementary tables}
\setcounter{table}{0} 
\begin{table*}[htp]
\small
\caption{\ Steady shear viscosity: fitting parameters for PL (Eq.~\ref{EqPL}) and Ellis models (Eq.~\ref{Eq(3)}).}
\label{Fitting}
\begin{tabular*}{\textwidth}{@{\extracolsep{\fill}}ccccccc}
\hline
Liquid sample & K (Pa.s$^n$) & $n$ (-) & $\eta_0$ (Pa.s) & $\tau_0$ (Pa) & $\alpha$ (-) & $\text{R}^2$ \\
\hline
1.19 \% w/w TUC & 1.82 & 0.28 & - & - & - & 0.99 \\
0.3 \% w/w cereal extract & - & - & 0.62 & 0.43 & 2.24 & 0.98\\
1 \% w/w cereal extract & - & - & 5.68 & 7.60 & 3.90 & 0.99\\
2\% w/w PEO & - & - & 0.39 & 23.99 & 2.24 & 0.96 \\
3\% w/w PEO & - & - & 3.22 & 22.83 & 2.07 & 0.99 \\
4\% w/w PEO & - & - & 10.37 & 27.11 & 2.08 & 0.99 \\
5\% w/w PEO & - & - & 31.12 & 35.32 & 2.11 & 0.98 \\
\hline
\end{tabular*}
\end{table*}

\end{document}